# The Capacity of the Interference Channel with a Cognitive Relay in Very Strong Interference

Stefano Rini†, Daniela Tuninetti*, Natasha Devroye* and Andrea Goldsmith†,
*University of Illinois at Chicago, Chicago, IL 60607, USA, Email: `danielat, devroye@uic.edu`
† Stanford University, Stanford, CA 94305, USA, Email: `stefano, andrea@wsl.stanford.edu`

*Abstract*—The interference channel with a cognitive relay consists of a classical interference channel with two source-destination pairs and with an additional cognitive relay that has *a priori knowledge* of the sources' messages and aids in the sources' transmission. We derive a new outer bound for this channel using an argument originally devised for the "more capable" broadcast channel, and show the achievability of the proposed outer bound in the "very strong interference" regime, a class of channels where there is no loss in optimality if both destinations decode both messages. This result is analogous to the "very strong interference" capacity result for the classical interference channel and for the cognitive interference channel, and is the first capacity known capacity result for the general interference channel with a cognitive relay.

## I. INTRODUCTION

Cognition is a rapidly emerging new paradigm in wireless communication whereby a node changes its communication scheme to efficiently share the spectrum with other users in the network. Cooperation among smart and well-connected wireless devices has been recognized as a key factor in improving the spectrum utilization and throughput of wireless networks [1]. The information theoretic study of cognitive networks has focused mostly on the cognitive interference channel, a variation of the classical interference channel where one of the transmitters has *perfect, a priori knowledge* of both the messages to be transmitted. Albeit idealistic, this form of *genie-aided* cognition has provided precious insights on the rate advantages that can be obtained with transmitter cooperation with one cognitive encoder. In this paper we study a natural extension of the cognitive interference channel where the genie-aided cognition, instead of being provided to only one of the users of the interference channel, is rather provided to a third node, a *cognitive relay*, that aids the communication between both source-destination pairs.

**Past work.** Few results are available for the InterFerence Channel with a Cognitive Relay (IFC-CR) and the fully general information theoretic capacity of this channel remains an open problem. The IFC-CR was initially considered in [2] where the first achievable rate region was proposed, and was improved upon in [3], which also provided a sum-rate outer bound for the Gaussian channel. This outer bound is based on an outer bound for the MIMO Gaussian cognitive interference channel and, in general, has no closed form expression. In [4] an achievable rate region was derived that contains all previously known achievable rate regions[1]. The first outer bounds for a general (i.e., not Gaussian) IFC-CR were derived in [5] by using the fact that the capacity region only depends on the conditional marginal distribution of the channel outputs. The authors of [5] first derived an outer bound valid for any IFC-CR and successively tighten the bound for a class of semi-deterministic channels in the spirit of [6], [7]. In the same paper, the tightened bound was also shown to be capacity for a the high-SNR binary linear deterministic approximation of the Gaussian channel, a model originally proposed in [8] for the classical IFC, for the case where the sources do not interfere at the non-intended destinations. In [9], with the insights gained from the high-SNR binary linear deterministic channel, we showed capacity to within 3 bits/sec/Hz for any finite SNR.

**Contributions.** In this paper we determine:

1) **a new outer bound for the interference channel with a cognitive relay** inspired by an argument originally devised for the "more capable" broadcast channel [10], also utilized in deriving the capacity of the cognitive interference channel in "weak interference" [11].
2) **a new outer bound in the "strong interference" regime** which is defined as the regime where-loosely speaking-the non-intended destination can decode more information than the intended destination. This regime parallels the "strong interference" regime for the interference [12] and the cognitive interference channels [13].
3) **capacity for the "very strong interference" regime**, that is, the regime where both decoders can, without rate loss, decode both messages. In this regime, the "strong interference" outer bound can be achieved with a simple superposition coding scheme.

**Paper Organization.** The rest of the paper is organized as follows: in Section II we formally introduce the channel model. In Section III we present a new outer bound for the general channel and an outer bound for the "strong interference" regime. In Section IV we show the achievability of the "strong interference" outer bound in the "very strong interference" regime. Section V specializes the results of the paper to the Gaussian interference channel with a cognitive relay. Section VI concludes the paper.

---
[1]The authors of [4] refer to the IFC-CR as "broadcast channel with cognitive relays", arguing that the model can also be obtained by adding two partially cognitive relays to a broadcast channel.

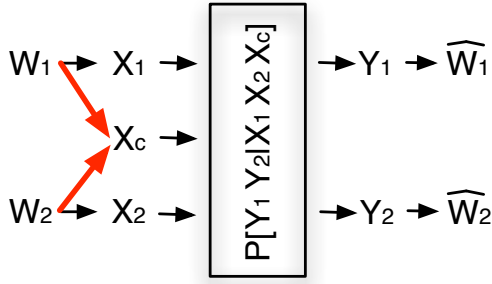

Fig. 1. The general memoryless IFC-CR channel model with two messages $W_1$ and $W_2$ known non-causally at the Cognitive Relay.

## II. CHANNEL MODEL

We consider the IFC-CR depicted in Fig. 1, in which the transmission of the two independent messages $W_i \in [1 : 2^{NR_i}]$, $i \in \{1,2\}$, is aided by a single *cognitive* relay, whose input to the channel has subscript $c$. The memoryless channel has transition probability $P_{Y_1,Y_2|X_1,X_2,X_c}$. A rate pair is achievable if there exists a sequence of encoding functions

$$X_1^N = X_1^N(W_1), \ X_2^N = X_2^N(W_2), \ X_c^N = X_c^N(W_1, W_2),$$

and a sequence of decoding functions

$$\widehat{W}_1 = \widehat{W}_1(Y_1^N), \ \widehat{W}_2 = \widehat{W}_2(Y_2^N),$$

such that

$$\lim_{N \to \infty} \max_{i=1,2} \Pr\left[\widehat{W}_i \neq W_i\right] = 0.$$

The capacity region is defined as the closure of the region of all achievable $(R_1, R_2)$-pairs. The capacity of the general IFC-CR is open. The IFC-CR subsumes three well-studied channels as special cases: (a) InterFerence Channel (IFC): if $X_c = \emptyset$; (b) Broadcast Channel (BC): if $X_1 = X_2 = \emptyset$; and (c) Cognitive InterFerence channel (CIFC): if $X_1 = \emptyset$ or $X_2 = \emptyset$.

## III. OUTER BOUNDS

The previously proposed an outer bound for the general memoryless IFC-CR in [5, Th.3.1] equals capacity when the channel reduces to a Gaussian CIFC in the "weak interference" [11, Lem.3.6], in the "very strong interference" regime [13, Th.6] and in the "primary decodes cognitive" regime [14, Th.3.1]. However, it does not reduce to the outer bound in [11, Th. 3.2], which is capacity for the CIFC in the "very weak interference" regime [11, Th.3.4], and for the semi-deterministic CIFC [15, Th.8.1]. For this reason we next derive a new outer bound inspired by the capacity of the "more capable" BC of [10] which does correspond to the outer bound of [11, Th.3.2] when the IFC-CR reduces to a CIFC. We also derive a simple expression from this first outer bound for a specific class of channels: the "strong interference" regime, where one of the users can more favorably decode the message of the other user that the intended receiver itself. This regime parallels the "strong interference" regime for the IFC [12] and the CIFC [13, Th.6].

**Theorem 1. "More capable" broadcast channel type outer bound.** If $(R_1, R_2)$ lies in the capacity region of the IFC-CR, then the following must hold:

$$R_1 \leq I(Y_1; X_1, X_c | X_2, Q), \tag{1a}$$
$$R_1 \leq I(Y_1; U_2, X_1 | Q), \tag{1b}$$
$$R_2 \leq I(Y_2; X_2, X_c | X_1, Q), \tag{1c}$$
$$R_2 \leq I(Y_2; U_1, X_2 | Q), \tag{1d}$$
$$R_1 + R_2 \leq I(Y_1; X_1, X_c | U_1, X_2, Q) + I(Y_2; U_1, X_2 | Q), \tag{1e}$$
$$R_1 + R_2 \leq I(Y_2; X_2, X_c | U_2, X_1, Q) + I(Y_1; U_2, X_1 | Q), \tag{1f}$$
$$R_1 + R_2 \leq I(Y_1; U_1 | Q) + I(Y_2; U_2 | Q), \tag{1g}$$
$$R_1 + R_2 \leq I(Y_1; X_1, X_2, X_c | Q) + I(Y_2; X_2, X_c | Y_1, X_1, Q), \tag{1h}$$
$$R_1 + R_2 \leq I(Y_2; X_1, X_2, X_c | Q) + I(Y_1; X_1, X_c | Y_2, X_2, Q), \tag{1i}$$

*for some input distribution* $P_{Q,X_1,X_2,X_c,U_1,U_2}$ *that factor as:*

$$P_Q P_{X_1|Q} P_{X_2|Q} P_{X_c|X_1,X_2,Q} P_{U_1,U_2|X_1,X_2,X_c,Q}. \tag{2}$$

*Proof:* The single-rate bounds in (1a) and (1c), as well as the sum-rate bounds in (1h) and (1i), were originally derived in [5, Th. 3.1]. The bound of (1d) is obtained as follows:

$$N(R_2 - \epsilon_N) \leq I(Y_2^N; W_2)$$
$$\stackrel{(a)}{\leq} \sum_{i=1}^{N} H(Y_{2,i}|Y_{2,i+1}^N) - H(Y_{2,i}|Y_{2,i+1}^N, W_2, X_2^N, Y_1^{i-1})$$
$$\stackrel{(b)}{\leq} \sum_{i=1}^{N} I(Y_{2,i}; U_{i,1}, X_{2,i}),$$

where (a) follows from the "conditioning reduces entropy" [16] property and (b) from defining:

$$U_{1,i} = [Y_1^{i-1}, W_2, X_2^{i-1}, X_{2,i+1}^N, Y_{2,i+1}^N], \tag{3}$$

and letting $X^0 = X^{N+1} = \emptyset$. (1d) is obtained by considering the time-sharing Random Variable (RV) $Q$ uniformly distributed on the interval $[0 : N]$ and independent of everything else. For the sum-rate bound in (1e):

$$N(R_1 + R_2 - 2\epsilon_N) \leq I(Y_1^N; W_1|W_2) + I(Y_2^N; W_2)$$
$$\leq \sum_{i=1}^{N} I(Y_{1,i}; W_1, Y_{2,i+1}^N | Y_1^{i-1}, W_2, X_2^N) + I(Y_{2,i}; W_2, X_2^N, Y_{2,i+1}^N)$$
$$\leq \sum_{i=1}^{N} I(Y_{1,i}; Y_{2,i+1}^N | Y_1^{i-1}, W_2, X_2^N) - I(Y_{2,i}; Y_1^{i-1} | W_2, X_2^N, Y_{2,i+1}^N)$$
$$+ I(Y_{1,i}; W_1 | Y_1^{i-1}, W_2, X_2^N, Y_{2,i+1}^N) + I(Y_{2,i}; W_2, X_2^N, Y_{2,i+1}^N, Y_1^{i-1})$$
$$\stackrel{(c)}{=} \sum_{i=1}^{N} I(Y_{1,i}; W_1 | U_{i,1}, X_{2,i}) + I(Y_{2,i}; U_{i,1}, X_{2,i})$$
$$\leq \sum_{i=1}^{N} I(Y_{1,i}; X_{1i}, X_{ci} | U_{1,i}, X_{2i}) + I(Y_{2,i}; U_{1,i}, X_{2i}),$$

where (c) follows from Csiszár's sum identity [17] and the definition of $U_{1,i}$ in (3).

The bounds in (1b) and (1f) are obtained similarly to the bounds in (1d) and (1e), respectively, by swapping the role of the sources and by defining:

$$U_{2,i} = [Y_2^{i-1}, W_1, X_1^{i-1}, X_{1,i+1}^N, Y_{1,i+1}^N]. \tag{4}$$

Finally, the bound in (1g) is obtained as follows:

$$N(R_1 + R_2 - 2\epsilon) \leq I(Y_1^N; W_1) + I(Y_2^N; W_2)$$
$$\leq \sum_{i=1}^{N} H(Y_{1,i}) + H(Y_{2,i})$$
$$- H(Y_{1,i}|Y_2^{i-1}, Y_{1,i+1}^N, X_1^{i-1}, X_{1,i+1}^N, W_1)$$
$$- H(Y_{2,i}|Y_1^{i-1}, Y_{2,i+1}^N, X_2^{i-1}, X_{2,i+1}^N, W_2)$$
$$= \sum_{i=1}^{N} H(Y_{1,i}) - H(Y_{1,i}|U_{2,i}) + H(Y_{2,i}) - H(Y_{2,i}|U_{1,i}).$$

■

*Remark* 1. Th. 1 is the tightest known outer bound for a general IFC-CR and it reduces to the capacity region of the "more capable" BC when $X_1 = X_2 = \emptyset$ in which case (1b) and (1e) are tight. Th. 1 also reduces the outer bound of [11, Th. 3.2] when either $X_2 = \emptyset$ or $X_1 = \emptyset$ in which case (1b), (1d) and (1e) are tight. However, Th. 1 does not reduce to the capacity region of the class of deterministic IFCs studied in [8] and to the outer bound for the semi-deterministic IFC in [7] when $X_c = \emptyset$. The difficulty in deriving outer bounds for the IFC-CR that are tight when the IFC-CR reduces to an IFC is also noted in [5]. The authors of [5, Th. 3.2] are able to derive tight bounds in this scenario by imposing additional constraints on the effect of interference on the channel outputs.

**Theorem 2. "Strong interference" outer bound.** *If*

$$I(Y_2; X_2, X_c|X_1) \leq I(Y_1; X_2, X_c|X_1) \quad (5)$$

*for all distributions*

$$P_{X_1,X_2,X_c} = P_{X_1} P_{X_2} P_{X_c|X_1,X_2}, \quad (6)$$

*then, if $(R_1, R_2)$ lies in the capacity region of the IFC-CR, the following must hold:*

$$R_1 \leq I(Y_1; X_1, X_c|X_2, Q), \quad (7a)$$
$$R_2 \leq I(Y_2; X_2, X_c|X_1, Q), \quad (7b)$$
$$R_1 + R_2 \leq I(Y_1; X_1, X_2, X_c|Q), \quad (7c)$$

*for some distribution*

$$P_{Q,X_1,X_2,X_c} = P_Q P_{X_1|Q} P_{X_2|Q} P_{X_c|X_1,X_2,Q}. \quad (8)$$

*Proof:* Similarly to [18, Lem. 4] and [12, Lem. 1], if the condition in (5) holds for all distribution in (6), then

$$I(Y_2; X_2, X_c|X_1, U) \leq I(Y_1; X_2, X_c|X_1, U),$$

for all $P_{X_1,X_2,X_c,U} = P_{X_1} P_{X_2} P_{X_c|X_1,X_2} P_{U|X_1,X_2,X_c}$. From this, it follows that when condition (5) holds, we can upper bound the bound in (1f) as:

$$I(Y_1; U_2, X_1|Q) + I(Y_2; X_2, X_c|X_1, U_2, Q)$$
$$\leq I(Y_1; U_2, X_1|Q) + I(Y_1; X_2, X_c|X_1, U_2, Q)$$
$$\leq I(Y_1; X_1, X_2, X_c, U_2|Q)$$
$$= I(Y_1; X_1, X_2, X_c|Q),$$

where the last equality follows from the Markov chain $Y_1 - (X_1, X_2, X_c) - U_2$ which is readily established by using the memoryless property of the channel to write

$$P_{Y_1,Y_2,X_1,X_2,X_c U_2}$$
$$= P_{Y_1,Y_2|X_1,X_2,X_c} P_{X_1,X_2,X_c,U_2}.$$
$$= P_{Y_1,Y_2|X_1,X_2,X_c} P_{X_1,X_2,X_c} P_{U_2|X_1,X_2,X_c}.$$

.
■

*Remark* 2. Given the symmetry of the channel model, Th. 2 also holds when the role of the sources is reversed. Although not valid for a general IFC-CR, Th. 2 is expressed only as a function of the channel inputs and does not contain auxiliary RVs as in Th. 1.

*Remark* 3. When condition (5) holds, it also implies that

$$I(Y_2; X_2, X_c|X_1, Y_1) \leq I(Y_1; X_2, X_c|X_1, Y_1) = 0, \implies$$
$$I(Y_2; X_2, X_c|X_1, Y_1) = 0. \quad (9)$$

Given (9), sum rate bound (1h) coincides with (7c). The bound (7c) is derived in [5] using the fact that capacity region does not depend on the conditional joint distribution of the channel outputs but only on their conditional marginal distribution. This observation can used to tighten the genie aided bounds as done by Sato in [19] for the BC. As for the CIFC of [15], the sum rate bound derived using Csiszár's sum identity coincides with the bound derived using Sato's idea in the "strong interference" regime.

## IV. CAPACITY IN "VERY STRONG INTERFERENCE"

In this section we show the achievability of the outer bound of Th. 2 in the "very strong interference" regime (to be defined later), which is a subset of the "strong interference" regime defined by (5) . This result parallels the "very strong interference" capacity result for the IFC [12] and the CIFC [13], where, under the "very strong interference" condition, the channel reduces to a compound two-user multiple access channel. For this class of channels the interfering signal at each receiver can be decoded without imposing any rate penalty and successively stripped from the received signal. Since the interference can always be distinguished from the intended signal, there is no need to perform interference pre-coding at the cognitive relay. This greatly simplifies the achievable scheme required to match the outer bound in Th.2. We will show in fact that a simple superposition coding schemes achieves Th. 2.

**Theorem 3. Capacity in "very strong interference".** *If* (5) *holds together with*

$$I(Y_1; X_1, X_2, X_c) \leq I(Y_2; X_1, X_2, X_c) \quad (10)$$

*for all distribution in* (6)*, then the region in* (7) *is capacity.*

*Proof:* Under the assumption of the theorem, the region in (7) is an outer bound for the considered IFC-CR. The achievability of the outer bound the region in (7) can be shown by considering a transmission scheme that employs two common messages, $U_{1c}, U_{2c}$ for source 1 and source 2, respectively, that are encoded in the channel inputs according to the distributions $P_{X_1|U_{1c}}$, $P_{X_2|U_{2c}}$ and $P_{X_c|U_{1c},U_{2c}}$. This scheme achieves the region:

$$R_1 \leq I(Y_1; U_{1c}|U_{2c}, Q), \quad (11a)$$
$$R_2 \leq I(Y_2; U_{2c}|U_{1c}, Q), \quad (11b)$$
$$R_1 + R_2 \leq I(Y_1; U_{1c}, U_{2c}|Q), \quad (11c)$$
$$R_1 + R_2 \leq I(Y_2; U_{1c}, U_{2c}|Q), \quad (11d)$$

for some input distribution that factors as:

$$P_Q P_{U_{1c}|Q} P_{U_{2c}|Q} P_{X_1|U_{1c},Q} P_{X_2|U_{2c},Q} P_{X_c|U_{1c},U_{2c},Q}, \quad (12)$$

where $Q$ is a time-sharing random variable defined as in Th. 1. Let now $U_{1c} = X_1$, $U_{2c} = X_2$ and $X_c$ a deterministic function of $X_1, X_2$. Under the condition in (10) the bound in (11d) can be dropped from the region in (11) and the resulting region coincides with (10). ∎

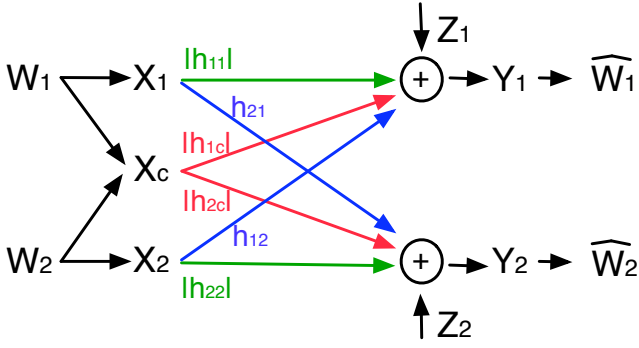

Fig. 2. The Gaussian IFC-CR in standard form.

## V. THE GAUSSIAN CASE

In the following we evaluate Th. 2 and Th. 3 for the Gaussian IFC-CR shown in Fig. 2. Without loss of generality (see App. A), we can restrict our attention to the Gaussian IFC-CR in *standard form* given by:

$$Y_1 = |h_{11}|X_1 + |h_{1c}|X_c + h_{12}X_2 + Z_1, \quad (13a)$$
$$Y_2 = |h_{22}|X_2 + |h_{2c}|X_c + h_{21}X_1 + Z_2, \quad (13b)$$

where $h_i \in \mathbb{C}$, $i \in \{11, 1c, 12, 22, 2c, 21\}$, are constant and known to all terminals, $Z_i \sim \mathcal{N}(0,1)$, $i \in \{1,2\}$, and $\mathbb{E}[|X_i|^2] \leq 1$, $i \in \{1,2,c\}$. The channel links $h_i, i \in \{11, 22, 1c, 2c\}$ can be taken to be real-valued without loss of generality because receivers and transmitters can compensate for the phase of the signals. The correlation among the noises is irrelevant because the capacity of the channel without receiver cooperation only depends on the noise marginal distributions.

**Theorem 4. The "strong interference" outer bound for the Gaussian IFC-CR.** *If*

$$\max_{|\beta_{2c}| \leq 1} ||h_{22}| + \beta_{2c}|h_{2c}||^2 - |h_{12} + \beta_{2c}|h_{1c}||^2 \leq 0 \quad (14)$$

$$(15)$$

*the capacity of a Gaussian IFC-CR is contained in the set:*

$$R_1 \leq \mathcal{C}\left(||h_{11}| + |h_{1c}|\beta_{1c}|^2\right), \quad (16a)$$
$$R_2 \leq \mathcal{C}\left(||h_{22}| + |h_{2c}|\beta_{2c}|^2\right), \quad (16b)$$
$$R_1 + R_2 \leq \mathcal{C}\left(||h_{11}| + |h_{1c}|\beta_{1c}|^2 + |h_{12} + |h_{1c}|\beta_{2c}|^2\right), \quad (16c)$$

*union over all* $(\beta_{1c}, \beta_{2c})$ *such that* $|\beta_{1c}|^2 + |\beta_{2c}|^2 = 1$.

*Proof:* Given the "Gaussian maximizes entropy" property [16] we have that the union over all the distributions in (8) of the region in (7) is equal to the union over all the zero-mean complex-valued proper Gaussian random vectors $[X_1, X_2, X_c]$ with covariance matrix

$$\begin{bmatrix} |\beta_{11}|^2 & 0 & \beta_{11}\beta_{1c} \\ 0 & |\beta_{22}|^2 & \beta_{22}\beta_{2c} \\ \beta_{11}^*\beta_{1c}^* & \beta_{22}^*\beta_{2c}^* & |\beta_{1c}|^2 + |\beta_{2c}|^2 + |\beta_{cc}|^2 \end{bmatrix} \quad (17)$$

for some

$$|\beta_{11}|^2 \leq 1, \quad |\beta_{22}|^2 \leq 1, \quad |\beta_{1c}|^2 + |\beta_{2c}|^2 + |\beta_{cc}|^2 \leq 1. \quad (18)$$

When considering the parametrization in (17) for the outer bound region in (7), we note that the choice

$$|\beta_{11}|^2 = |\beta_{22}|^2 = 1, \quad \beta_{cc} = 0, \quad |\beta_{1c}|^2 + |\beta_{2c}|^2 = 1, \quad (19)$$

yields the largest region. Since the region in (7) can be obtained from the union over the parameter set in (19) rather than the larger set in (18), condition (5) only needs to hold for all the complex Gaussian inputs satisfying (19). ∎

**Theorem 5. Capacity in "very strong interference" for the Gaussian IFC-CR.** *If, in addition to condition* (15), *we also have*

$$\max_{|\beta_{1c}|^2 + |\beta_{2c}|^2 = 1} ||\beta_{1c}|h_{1c}| + |h_{11}||^2 + |h_{12} + \beta_{2c}|h_{1c}||^2$$
$$- |h_{21} + |h_{2c}|\beta_{1c}|^2 - ||h_{22}| + |h_{2c}|\beta_{2c}|^2 \leq 0, \quad (20)$$

*the region of* (16) *is capacity.*

*Proof:* Since the outer bound in (16) is obtained as the union over all the zero-mean complex Gaussian inputs parameterized by (17) satisfying (19), inequality (10) needs to hold only for this choice of RVs. ∎

A representation of the "strong interference" regime of Th. 4 and the "very strong interference" regime of Th. 5 for the Gaussian IFC-CR is shown in Fig. 3. We focus on the case of positive, real channel coefficients and inputs with symmetric cognitive links, that is $|h_{1c}| = |h_{2c}| = |h_c|$. In this case the "strong interference" condition of (15) simplifies to $|h_{12}| \geq |h_{22}|$ and the "very strong interference" regime of (20) to

$\max_{|\beta_{1c}|^2+|\beta_{2c}|^2=1} 2h_c \left((|h_{12}| - |h_{22}|)\beta_{2c} + (|h_{11}| - |h_{21}|)\sin(x)\right)$
$+|h_{11}|^2 + |h_{12}|^2 - |h_{22}|^2 - |h_{21}|^2 \leq 0$
$\max_x 2h_c \left((|h_{12}| - |h_{22}|)\cos(x) + (|h_{11}| - |h_{21}|)\sin(x)\right)$
$+|h_{11}|^2 + |h_{12}|^2 - |h_{22}|^2 - |h_{21}|^2 \leq 0$

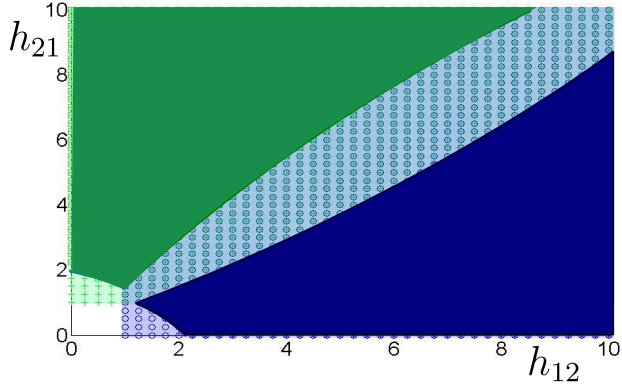

Fig. 3. The "strong interference" regime of Th. 4 (light blue) and the "very strong interference" regime of Th. 5 for user 1 as well as the 'strong interference" regime (light green) and the "very strong interference" regime (dark green) for user 2 for the Gaussian IFC-CR with $h_{11} = h_{22} = h_{1c} = h_{2c} = 1$ on the plane $h_{12} \times h_{21} = [0:10] \times [0:10]$.

## VI. CONCLUSION AND FUTURE WORK

We introduce a new outer bound for the interference channel with a cognitive relay and show the achievability of this outer bound in the "very strong interference" regime by having both decoders decode both messages as in a compound multiple access channel. This result is very similar in nature to the "very strong interference" capacity results for the interference channel and the cognitive interference channel. Although significant, the contributions of this paper are only the first step to a better understanding of the capacity region of the cognitive interference channel with a cognitive relay which remains largely undiscovered.

## ACKNOWLEDGMENT

The work of the D. Tuninetti and N. Devroye was partially funded by NSF under awards number 0643954 and 1017436. The contents of this article are solely the responsibility of the authors and do not necessarily represent the official views of the NSF.

## APPENDIX A
### THE IFC-CR IN STANDARD FORM

A general IFC-CR is expressed as

$$\widetilde{Y}_1 = \widetilde{h}_{11}\widetilde{X}_1 + \widetilde{h}_{1c}\widetilde{X}_c + \widetilde{h}_{12}\widetilde{X}_2 + \widetilde{Z}_1, \quad (21a)$$
$$\widetilde{Y}_2 = \widetilde{h}_{22}\widetilde{X}_1 + \widetilde{h}_{2c}\widetilde{X}_c + \widetilde{h}_{21}\widetilde{X}_1 + \widetilde{Z}_2, \quad (21b)$$

for $\widetilde{h}_i$ $i \in \{11, 22, 1c, 2c, 12, 21\}$, $E[|\widetilde{X}_j|^2] \leq \widetilde{P}_j$ $j \in \{1, 2, c\}$ and $E[|\widetilde{Z}_k|^2] = \sigma_k^2$ $k \in \{1, 2\}$. Consider now the transformation

$$Y_1 = \frac{\widetilde{Y}_1}{\sigma_1} e^{-j\angle \widetilde{h}_{1c}} \qquad Y_2 = \frac{\widetilde{Y}_2}{\sigma_2} e^{-j\angle \widetilde{h}_{2c}}$$
$$X_1 = \frac{\widetilde{X}_1}{\sqrt{\widetilde{P}_1}} e^{-j(\angle \widetilde{h}_{11} + \angle \widetilde{h}_{1c})} \qquad X_2 = \frac{\widetilde{X}_2}{\sqrt{\widetilde{P}_2}} e^{-j(\angle \widetilde{h}_{22} + \angle \widetilde{h}_{2c})}$$
$$X_c = \frac{\widetilde{X}_c}{\sqrt{\widetilde{P}_c}}$$
$$|h_{11}| = \frac{\sqrt{\widetilde{P}_1}|\widetilde{h}_{11}|}{\sigma_1} \qquad |h_{22}| = \frac{\sqrt{\widetilde{P}_2}|\widetilde{h}_{22}|}{\sigma_2}$$
$$|h_{1c}| = \frac{\sqrt{\widetilde{P}_c}|\widetilde{h}_{1c}|}{\sigma_1} \qquad |h_{2c}| = \frac{\sqrt{\widetilde{P}_c}|\widetilde{h}_{2c}|}{\sigma_2}$$
$$|h_{12}| = \frac{\sqrt{\widetilde{P}_2}\widetilde{h}_{12}}{\sigma_1} e^{-j\angle \widetilde{h}_{11}} \qquad |h_{21}| = \frac{\sqrt{\widetilde{P}_1}\widetilde{h}_{21}}{\sigma_2} e^{-j\angle \widetilde{h}_{22}}, \quad (22)$$

since the transformation in (22) is a linear transformation, the channel in (21) is equivalent to the channel in (13).